\documentclass[]{elsarticle}

\usepackage{lineno,hyperref}
\usepackage{graphicx}
\usepackage{amsmath}
\usepackage{booktabs}       

\newtheorem{mydef}{Definition}

\DeclareMathOperator{\argmin}{arg\,min}

\begin{document}

\begin{frontmatter}

\title{Generating behavioral acts of predetermined apparent complexity}

\author{A.V.~Olifer}
\address{School of Science and Technology, Georgia Gwinnett College, 1000 University Center Lane, Lawrenceville, GA 30043, USA}
\ead{aolifer@ggc.edu}

\begin{abstract}
 Behavior of natural and artificial agents consists of  behavioral episodes or acts.  This study introduces a quantitative measure of  behavioral acts -- their apparent complexity. The measure is based on the concept of the Kolmogorov complexity. It is an apparent measure because it is determined solely by the readings of the signals that directly encode percepts and actions during behavior. The article describes an algorithm of generating behavioral acts of predetermined apparent complexity. Such acts can be used to evaluate and develop learning abilities of artificial agents.   
\end{abstract}

\end{frontmatter}


\section{Introduction}
\label{Introduction}
Achievements of artificial intelligence often relate to specific tasks, for example winning the game of Go (AlphaGo, Google DeepMind) or interacting with people (robot Sophia, Hanson Robotics). The next step could be the agents able to learn tasks of any nature provided the behavior required for attaining the task is possible with the sensors and actuators the agents have. For example, Sophia's descendants may not just talk but also play Go by moving stones on the board. 

To develop such agents it might be beneficial to have achievable tasks of various complexity so that agents could gradually advance in learning. A related goal is to create measures and tests for evaluating and comparing agents \cite{Hu01,HO00,LV13,Sch02}. Known tasks often stimulate designing agents for task-specific behaviors (reviewed in \cite{HO15}). The complexity of these tasks is defined in task-specific ways as well (finding goal from one direction, etc. \cite{FM00}). Such measures of complexity cannot be applied to the tasks of different nature. 

\begin{figure}[h]
	\includegraphics[width=0.9\linewidth]{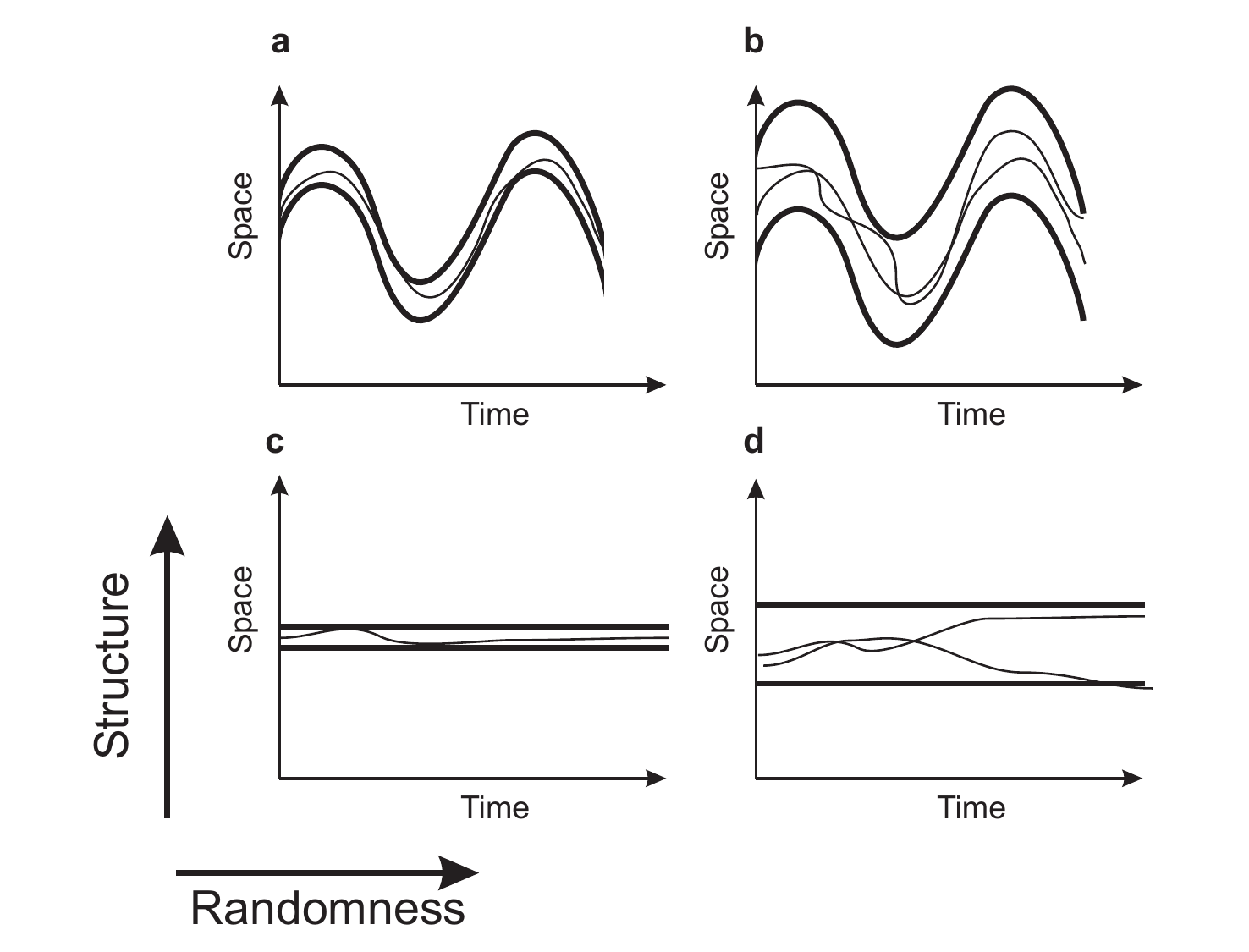} 
	\caption{Structure and randomness of behavior. Schematic representation of how two fundamental characteristics of sets of objects, structure and randomness, distinguish behavioral acts. Behavior (a) has a relatively complex structure and allows only small random distortions. Behavior (b) has a relatively complex structure but allows larger distortions than in (a). Behavior (c) has a simple structure and allows only small distortions.	  Behavior (d) has a simple structure and allows large random distortions. "Space" is an abstract space in which behaviors are defined; thick lines - borders of feasible behaviors, thin lines - instances of behaviors.}
	\label{SandR}
\end{figure}

Complex tasks are associated with complex acts of behavior required to accomplish the tasks' goals. The present study focuses on the apparent complexity of behavioral acts. By apparent complexity we understand the complexity of readings of the agent's sensors and actuators. Intuitively, the behavioral act is complex if the time series of the readings that are associated with the act  1) have few or no patterns, 2) are long, and 3) cannot vary or can vary only a little from one instance of the act to another, that is the behavior needs to be accurate. These three properties reflect the structure and randomness of a behavioral act (Fig.\ref{SandR}). This general view (see, e.g., \cite{TAO08}) is put into work here using the Kolmogorov complexity. 

The Kolmogorov complexity of an object, in this case a behavioral act, is essentially the length of its smallest possible description \cite{LV08}. The concept has two issues related to the present subject. The first one is the incomputability of the Kolmogorov complexity. It means that in general the Kolmogorov complexity cannot be calculated exactly. However there are computable approximations of the Kolmogorov complexity that proved to be useful in various applied fields \cite{CV05}. The second issue is that real-world individual behavioral acts have  variability of sensory inputs and actions. To take this into account one needs to consider the Kolmogorov complexity not of one instance of the behavioral act but of the set of all possible instances.

\vspace{\baselineskip}
\vspace{\baselineskip}

Accordingly, two main assumptions are made here. First, that the readings of the sensors and actuators encoding rewards, percepts, and actions are fully known for every behavioral instance. Second, that the set of feasible behavioral instances of the act is either known or can be approximated. Explicit agent-environment interactions \cite{HO15,LV13} as well as the processes of learning and behavioral control are not considered in the present study.  

The complexity of time series has been intensely studied using Information Theory methods \cite{B01,P15}.
One approach is to accumulate statistics of a behavior and then use the Information Theory to quantify how predictive the behavior is \cite{B01,P15}. Simple behaviors are easily predicted and vise versa.
The main difference between the proposed approach and Information Theory methods is in the focus of analysis. Information Theory methods strive to quantify the complexity of the process or the model generating a set of time series. The Kolmogorov complexity evaluates how hard is to compute or describe an individual series.

In this not we describe how to evaluate the complexity of behavioral acts and generate acts of predetermined complexity. The provided algorithm and examples show that acts of predetermined complexity can be generated using moderate computational resources. 

\section{Background}
We define a behavioral instance as a triple of time series $(R,P,A)$ of rewards, percepts, and actions of the agent. In animals, percepts are represented by the activity of all types of sensors, including those responsible for homeostasis. Actions are represented by the activity of all neurons that control muscles and glands. In artificial agents, rewards, percepts, and actions are represented by the readings of the corresponding sensors and actuators.  

Each of the three components of the act has its own complexity. For example, complex percepts may require simple actions, and so on. The complexity of each component is determined by the complexity of that component in a typical instance of the act and by the number of possible variations of that component in the feasible behavioral instances.
For example, if the component in a typical instance is simple, and multiple variations in the feasilble instances are possible -- the complexity of this component of the act is intuitively small (Fig.\ref{SandR}d). On the contrary, if the component in a typical instance is complex, and only few variations of that component are possible -- the complexity of this component of the act is intuitively large (Fig.\ref{SandR}a).

We formalize this intuition using the concept of the Kolmogorov complexity. A fundamental property of the Kolmogorov complexity is that it is a characteristic of an individual object without a reference to the distribution from which this object comes from. Informally, the Kolmogorov complexity $K(x)$ of an object $x$ is the length, in bits, of the shortest program that generates a full description of $x$ and then halts. A rigorous definition,  along with the proofs of the properties of the Kolmogorov complexity considered below, can be found in the encyclopedic treatise by Li and Vit{\'a}nyi \cite{LV08}. 

The Kolmogorov complexity is not computable \cite{LV08}.  Analytical estimates of the Kolmogorov complexity often employ the fact that if the elements of a set $S$ are indexed then the index of an element has the length of at most $\lceil log \, |S|\rceil$ bits. Here $|S|$ is the cardinality of $S$, the logarithm as all the logarithms onwards is base two, and $\lceil \cdot \rceil$ is the ceiling function. In estimates of the Kolmogorov complexity, the error introduced by omitting the ceiling function is ignored.  

To obtain a numerical estimate of the Kolmogorov complexity of a bitstring the file with the string is compressed. The  length of the compressed file is taken as an estimate of the Kolmogorov complexity. This study uses gzip archiver for compressing. At the initial stage we also used bzip2 and got similar results in agreement with  observations of others \cite{CV05}. 

The following examples illustrate strengths and weaknesses of using gzip to approximate the Kolmogorov complexity. 

{\emph{Example 1:}} A 1024 byte--long random bitstring after compression by gzip became 1053 bytes long. The result of 1053 bytes was the same for several files with independent random bitstrings. The Kolmogorov complexity of a random bitstring cannot be less than the length of the string since there are no patterns to compress. The file length after compression increased compared to the original length since gzip needs extra bytes, in particular an eight-byte footer with a checksum. 

{\emph{Example 2:}} A 1024 byte--long file of zeros in every bit was compressed to 35 bytes. 35 bytes is longer than the two bytes needed to represent $2^{13}$, the number of bits in 1024 bytes, yet the level  of compression is quite high because of the apparent simplicity of the bitstring of zeros.

{\emph{Example 3:}} Let $b_i$ and $b_j$, $i<j$, be two successive bits equal to one, $b_i=1$, $b_j=1$, with the bits in between equal to zero, $b_k=0$, $k=i+1,\cdots,j-1$. Transform the original string so that $b_i=1$, $b_j=1$, but $b_k=1$, $k=i+1,\cdots,j-1$. In other words bits equal to one in the first string are considered as the endpoints of the intervals of the bits equal to one in the transformed string. For example, '010010..' transforms into '011110..' and so on. 
This transformation does not change the length of compressed files considerably. For example, a file of 1024 bytes with 1310 random bits equal to one was compressed to 769 bytes. The second file, with the transformed bitstring, was compressed to 735 bytes. The lengths of the files with bitstrings transformed in this way never differed for more than 5\%.

\section{Apparent complexity of behavioral acts}
\label{C_actions}
In what follows we define how to evaluate the complexity of actions. The other two components of behavioral acts, perceptions and rewards, are evaluated in the same way.

Let $\mathcal{A}$ be the space in which each dimension represents the activity of a certain actuator. Suppose $\mathcal{A}$ is properly discretized and bijectively mapped to $\{0,1\}^{n_A}$, the set of all bitstrings of length $n_A$. Without loss of generality it is further assumed that $\mathcal{A}=\{0,1\}^{n_A}$. Time is discretized as well. At any moment of time, the state of all the actuators is represented by a point in $\mathcal{A}$.  

Suppose a behavioral act lasts $T$ time steps and $M$ instances of it have been observed. Let $A_{obs}=\{a^1,a^2,\cdots,a^M\}$ be the set of the corresponding actions. Each instance $a^j$ is a bitstring of the length ${n_A}\times (T+1)$. Next, let $A$ denote the set of all possible feasible actions associated with the act, $A_{obs}\subseteq A$; if the act requires accurate movements then $|A|$, the cardinality of $A$, should be small. Finally, let $K(a)$ be the Kolmogorov complexity of $a \in A$.

The following definition reflects the intuition that low accuracy of actions decreases the complexity of actions associated with the act and vice versa.

\vspace{0.2in}
\begin{mydef} 
	\label{Def1}
	The complexity of actions $A$ associated with an act is
	\begin{equation}
	\label{C_A}
	C(A) = max\,(mean_{\, a\in A}\, K(a)-log\,|A|,0).
	\end{equation}
	\noindent
\end{mydef}

\vspace{\baselineskip}
\vspace{\baselineskip}

This definition follows from a fundamental equality \cite[p.~403]{LV08},
$$K(S) + log\,|S|=K(x)+O(1),$$

\noindent
showing  that a minimal description of a typical representative $x$ of a set $S$  up to a constant that does not depend on $x$  consists of the minimal description of $S$ and the minimal description of $x$ in $S$; the latter does not exceed $log\,|S|$. Maximization in (\ref{C_A}) guarantees non-negativity of $C(A)$. According to (\ref{C_A}), sampled instances allow only for an estimate of $C(A)$. 

Definition \ref{Def1} bears similarity with the definition of the complexity based on Predictive Information. Both contrast randomness and predictability \cite{B01}. We hypothesize both definitions are equivalent in an asymptotic sense.

The following example illustrates an application of (\ref{C_A}).

{\emph{Example 4:}} Consider two sets of 1024 byte-long bitstrings. In the bitstrings of the first set, $S1$, the first four bits in every byte  are filled randomly with zeros and ones while the last four bits are always set to ones. An example would be '01101111 11011111 ...' .   In the bitstrings of the second set, $S2$, the first byte is filled with ones and zeros exactly as in set $S1$. However all subsequent byte are not random. In each of them the first four bits are obtained from the first four bits of the previous byte by a cyclic permutation. As in $S1$, the last four bits are ones in every byte.  For example, if the first byte was '00111111' then the second byte would be '01101111', the third one '11001111', and so on.  

\begin{table}
	\vspace{\baselineskip}
	\caption{The compressed length, in bytes, of two types of files. The files of the first type, "Random bits", had random values of the first four bits in every byte. The files of the second type, "Periodic bits", had repeated first four bits with the period of four bytes. The last four bits in every byte were equal to one in all the files. Before compression, every file had the length of 1024 bytes. The lengths after compression:  mean = 519.9 bytes, std = 3.5 bytes for "Random bits"; mean = 46.9 bytes, std = 1.6 bytes for "Periodic bits"; n = 10.}  
	\vspace{\baselineskip}
	\begin{center}
		\begin{tabular}{l *{10}{c}}
			\toprule
			File & \#1 & \#2 & \#3 & \#4 & \#5 & \#6 & \#7 & \#8 & \#9 & \#10 \\
			\midrule
			Random bits & 521 & 522 & 525 & 517 & 521 & 520 & 519 & 520 & 522 & 512 \\
			\midrule
			Periodic bits & 42 &	45 & 42 & 44 & 41 &	44 & 44 & 44 & 44 &	44 \\
			\bottomrule
		\end{tabular}
	\end{center}
\end{table}
\vspace{\baselineskip}

Suppose sets $S1$ and $S2$ correspond to the actions of the agent during two different behavioral acts. Which act requires more complex actions?
Table 1 shows the lengths (in bytes) of ten randomly generated strings of $S1$ and $S2$ after gzip compression. These lengths allow for calculating the first entity needed in (\ref{C_A}), the mean complexity of an instance. For $S1$ the mean complexity was equal to $4159$ bits $(4159 = 519.9 \times 8)$. For $S2$, the mean complexity was equal to $347$ bits $(347 = 43.4 \times 8)$. This shows that the elements of $S1$ were on average more than ten times complex than the elements of $S2$. But what are the cardinalities of these sets? In the case of $S1$ there are 16 possible different combinations of 4 bits in each of 1024 bytes. It gives the total of $2^{4\cdot 1024}$ possible instances. Thus $|S1|=2^{4096}$, $log\, |S1| = 4096$ bits, and according to (\ref{C_A}), $C(S1) = 4159 - 4096 = 63$ bits. In contrast, $|S2|$ is small. It has only 16 instances defined by the possible combinations of the first four bits. Hence $log \, |S2| = 4$ bits, and  $C(S2) = 347 - 4 = 343$ bits, which is about 5.5 times more than $C(S1)$. Thus the second behavioral act, $S2$, is more complex compared to $S1$. The "structural" component of $S2$ is relatively simple because of the periodicity. Its "randomness" is low since only small variability in actions is allowed. 

The complexity of rewards and percepts of behavioral acts can be evaluated in the same way as the complexity of actions. Together, they determine  the complexity of the behavioral act. 

Let $R$, $P$, and $A$ be subsets of the reward space $\mathcal{R}$, percept space $\mathcal{P}$, and action space $\mathcal{A}$ respectively. 

\begin{mydef}
	\label{BAC}
	The complexity of a behavioral act $(R,P,A)$ is the triple $(C(R),C(P),C(A))$, where $C(R)$, $C(P)$, and $C(A)$ are the Kolmogorov complexities of the sets $R$, $P$, and $A$ defined according to (\ref{C_A}).
\end{mydef}

To use (\ref{C_A}) in the case of observed behaviors two estimates are needed for each of the three components: $R$, $P$, and $A$. First, it is necessary to evaluate the complexity of typical instances. Second, it is necessary to evaluate the numbers of feasible instances. Underestimating the number of different feasible instances leads to overestimating the complexity; see (\ref{C_A}). 

\section{Actions of predetermined complexity}
\label{Generating}

This section describes how to generate components of behavioral acts with a predetermined complexity. Actions are considered for specificity. The other components are generated in the same way.

According to Definition \ref{Def1}, the complexity of actions is defined by two components: the complexity of a typical time series of actions and the cardinality of the set of all possible actions consistent with the act. The cardinality is assumed to be known. Then a sufficient complexity $K$ of a typical time series follows from (\ref{C_A}).  First we show how to generate a time series with the complexity equal to $K$; it must be understood that in this and other cases the Kolmogorov complexity of an object can be determined only approximately. The value of the cardinality is then used to define possible distortions of the generated time series. 

A bitstring of predetermined complexity can be obtained by assigning a certain number of randomly selected bits to one while keeping all the other bits equal to zero. Let $L$ be the number of bits equal to one in a bitstring $s$ of length $n$. There are $\binom{n}{L}$ of such strings. To describe $s$ it is suffice to describe the index of $s$ in the set of $\binom{n}{L}$ possible indexes. The description of the index takes no more than $log\,\binom{n}{L}$ bits. The value of $L$ is  determined from the requirement that the description length is the smallest among those exceeding $K$: 

\begin{equation}
\label{basic_eqn_L}
L = \argmin_{L'}\{log\,\binom{n}{L'}\geq K\}.
\end{equation}    

First of all, the inequality in (\ref{basic_eqn_L}) shows that to be complex, the string has to be long enough. Indeed, the inequality has no solutions for $n$ such that $\binom{n}{n/2}\leq 2^K$. If $L$ satisfies (\ref{basic_eqn_L}) then $n-L$ also satisfies (\ref{basic_eqn_L}). This symmetry reflects the fact that when all the zeros and ones in string $s$ are flipped, the complexity of $s$ does not change.  

In practice, one can find $L$ from (\ref{basic_eqn_L}) using the bisection method. Binomial coefficients increase monotonically when $L'$ changes from $0$ to $n/2$. The method will require at most $log(n)$ iterations.

Asymptotically,  inequality  (\ref{basic_eqn_L}) turns into an inequality with an entropy. Namely, suppose $n$ is large and  $L\sim \lambda n$ and $K\sim \kappa n$ for large $n$; here $L$ is the solution to  (\ref{basic_eqn_L}) .   Then  the approximation of the factorials in the binomial coefficient in (\ref{vectorL}) using Stirling's formula $n!\sim \sqrt{2\pi n}n^ne^{-n}$ yields

\begin{equation}
\label{full_entropy_simple}
H(\lambda)\geq \kappa,
\end{equation}

\noindent
where $H(\lambda)= -\lambda log\,\lambda - (1-\lambda)log\,(1-\lambda)\}$.    A minimal $\lambda$ that satisfies (\ref{full_entropy_simple}) can be used an an (initial) approximation to $L$. 

The same approach can be used for time series of strings. If there are no limitations on the number of bits equal to one at every time step, then all $s_i$ can be concatenated into one string and for that string the number of bits equal to one can be determined according to (\ref{basic_eqn_L}).

\begin{equation}
\label{basic_eqn_L}
L = \argmin_{L'}\{log\,\binom{n(T+1)}{L'}\geq K\}.
\end{equation} 

Provided  $L\sim \lambda n(T+1)$ and $K\sim \kappa n(T+1)$ when $n$ is large the Stirling's formula leads to inequality (\ref{full_entropy_simple}) as well (just consider $n(T+1)$ instead of $n$).  If the number of ones $L$ at all of $T+1$  time step should be the same, then $L$ is defined using a multinomial coefficient:

\begin{equation}
\label{vectorL}
L=\argmin_L\{log\,\binom{n(T+1)}{L,n-L,\cdots,L,n-L}\geq K\}.
\end{equation}   

\noindent
On average, a  time series consisting of $T+1$ strings of length $n$ with $L$ bits equal to one in each of them is more complext compared to the string of length $n(T+1)$ with $L(T+1)$ bits  equal to one since

$$\binom{n(T+1)}{L(T+1)}<\binom{n(T+1)}{L,n-L,\cdots,L,n-L}.$$

In other words, on average the complexity of a string of length $n(T+1)$ bits can be achieved with a smaller number of bits equal to one if these bits are equaly spread between  $T+1$ substrings of length $n$.

Interestingly, however, for large values of $n$ the difference between the corresponding entropies vanishes. Indeed, assuming $L\sim \lambda n$ and $K\sim \kappa n(T+1)$ when $n$ is large and applying the Stirling formula to the multinomial the inequality (\ref{vectorL}) leads to the same inequality (\ref{full_entropy_simple}) for $\lambda$. In other words, asymptotically, $\lambda$, the average probability that a randomly selected bit is equal to one, could be the same.

The above formulas can be directly generalized for the case in which the number of bits equal to one in each $s_i$ could be equal to $L_i$,  $0\leq L_i\leq n$, $i=0,\cdots, T$, and $L_i$ must be from some set $R$.  Then $L_i$ are found from

\begin{equation}
\label{vectorL_R}
(L_0,L_1,\cdots,L_{T+1})=\argmin_{(L'_0,L'_1,\cdots,L'_{T+1})\in R}\{log\,\binom{n}{L'_0,n-L'_0,\cdots,L'_{T+1},n-L'_{T+1}}\geq K\}.
\end{equation}   

If for large $n$, $L_i \sim \lambda_i n$, $0\leq \lambda_i\leq 1$, and $K=\kappa n(T+1)$ then the approximation of the multinomial coefficient in (\ref{vectorL}) using the Stirling's formula yields:

\begin{equation}
\label{full_entropy}
H(\lambda)\geq \kappa,
\end{equation}

\noindent
where $H(\lambda)= \sum_{i=0}^{T} H(\lambda_i) = \sum_{i=0}^{T} \{-\lambda_i log\,\lambda_i - (1-\lambda_i)log\,(1-\lambda_i)\}$.  

\vspace{\baselineskip}
Now everything is ready to describe an algorithm of generating a time series of the action $A$ of complexity $C(A)$. The generated time series will be playing the role of a typical (or average) action. Then, using $|A|$, the cardinality of $A$, one can determine possible distortions of that series.  Since this is going to be a "staged" behavioral act it is assumed that  $|A|$ is known.

\vspace{\baselineskip}
\hrule height 1.5pt
\vspace{0.075in}
{\bf{ALGORITHM}}
\vspace{0.075in}
\hrule 
\vspace{0.075in}
{\bf{Require:}} $C(A)$ -- the target complexity of the set of actions $A$, $|A|$ -- the cardinality of $A$, $n_A$ -- the number of bits at every time step,  $T$ -- the number of time steps, $R$ - the set of possible values of the number of ones. 

\quad 1. Calculate $K = C(A)+ log\,|A|$; cf. (\ref{C_A}). 

\quad 2. Check if $T$ and $n_A$ are sufficiently large. If the inequality (\ref{full_entropy}) holds

\qquad then CONTINUE, else STOP.

\quad 3. Determine the number of bits to be assigned to one, using formulas  (\ref{vectorL_R}),

\qquad (\ref{vectorL}), or (\ref{basic_eqn_L}).

\quad 4. Generate a sequence of strings of  complexity $K$ by assigning to one 

\qquad the number of  bits, obtained in 3., in random positions. Set other bits 

\qquad to zeros.
\vspace{0.05in}
\hrule height 1.5pt
\vspace{\baselineskip}

{\emph{Example 5:} In this example, the goal was to use the algorithm to generate time series that could be considered as typical action instances for the action  with the  complexity $C(A)=2048$ bits and the number of bits at every time step $n_A=8$. The number of bits equal to one, $n^1_A$, was determined for the whole time series. Bitstrings, representing typical instances of the actions, varied from trial to trial. The lengths of the corresponding compressed files varied as well. However, the variances of the lengths were small.
	
	For the first action, the input was $C(A)= 2048$ bits, $n_A=8$ bits, $T=255$, $|A|=0$. 
	Step 1 resulted in $K=2048$. In Step 2, the value of $T$ was found to satisfy $n_A(T+1)=K$. In Step 3, $n^1_A$ was found according to (\ref{basic_eqn_L}) to be equal to $n^1_A = 0.5n_A = 1024$ bits. In step 4, bits were set to one with probability $0.5$, which resulted in $1026$ bits equal to one (Fig.\ref{AAlg}a). The corresponding file was compressed to $2280$ bits. By construction ($|A|=0$) , distortions were not allowed for this particular action.
	
	For the second action, the input was $C(A)= 2048$ bits, $|A|=0$, $n_A=8$ bits, $T=1023$. Step 1 resulted in $K=2048$. In Step 2, the value of $T$ was found to satisfy $n_A(T+1)>K$. In Step 3, $n^1_A = 344$ bits was found according to (\ref{basic_eqn_L}). In step 4, bits were set to one with the probability $0.042=344/8192$, which resulted in $349$ bits equal to one (Fig.\ref{AAlg}b). The corresponding file was compressed to $3328$ bits. By construction, distortions were not allowed for this particular action.
	
	For the third action, the input was $C(A)= 2048$ bits, $|A|=2^{2048}$, $n_A=8$ bits, $T=1023$. Note that in this case distortions were allowed. Step 1 resulted in $K=4096$. In Step 2, the value of $T$ was found to satisfy $n_A(T+1)>K$. In Step 3, $n^1_A = 912$ bits was found according to (\ref{basic_eqn_L}). In step 4, bits were set to one with the probability $0.111=912/8192$, which resulted in $886$ bits equal to one (Fig.\ref{AAlg}c). The corresponding file was compressed to $5312$ bits.  
	
	How the cardinality $|A|$ of the set of all allowed distortions translates into the boundaries of $A$ depends on additional constraints to actions. In the simplest case, $A=B(N,\rho)$ is the ball of radius $\rho$ in the space $\{0,1\}^N$, $N=n(T+1)$, centered at the point represented by a typical action instance. To determine the radius of the ball one can use the Hamming distance. The Hamming distance between binary strings equals to the number of bits, in which the values of the two strings differ. In this case, if the radius $\rho = p N$, $0< p \leq 0.5$, then $|V|\leq 2^{H(p)N}$, where $H(p)$ is the binary Shannon's entropy $H(p)=-p\,log\,p-(1-p)\,log\,(1-p)$ \cite[p.~52]{CT91}. It is easy to find that $H(p)=0.25$ when $p\approx 0.04$. It implies that the ball has no more than $|A|=2^{2048}=2^{0.25*8192}$ elements if the radius does not exceed $327 =\lfloor 0.04*8*1024\rfloor$; here $\lfloor \cdot \rfloor$ is the floor operation. Hence, other actions corresponding to the case considered, can be at distance of at most $327$ bits from the sequence shown in (Fig.\ref{AAlg}c). These actions can be obtained by flipping at most $327$ randomly chosen bits in the sequence from (Fig.\ref{AAlg}c). Fig.\ref{AAlg}c relates to Fig.\ref{AAlg}b as Fig.\ref{SandR}b relates to Fig.\ref{SandR}a.

	\vspace{\baselineskip}
	\begin{figure}[h!]
		\begin{center} 
			\includegraphics[width=0.9\linewidth]{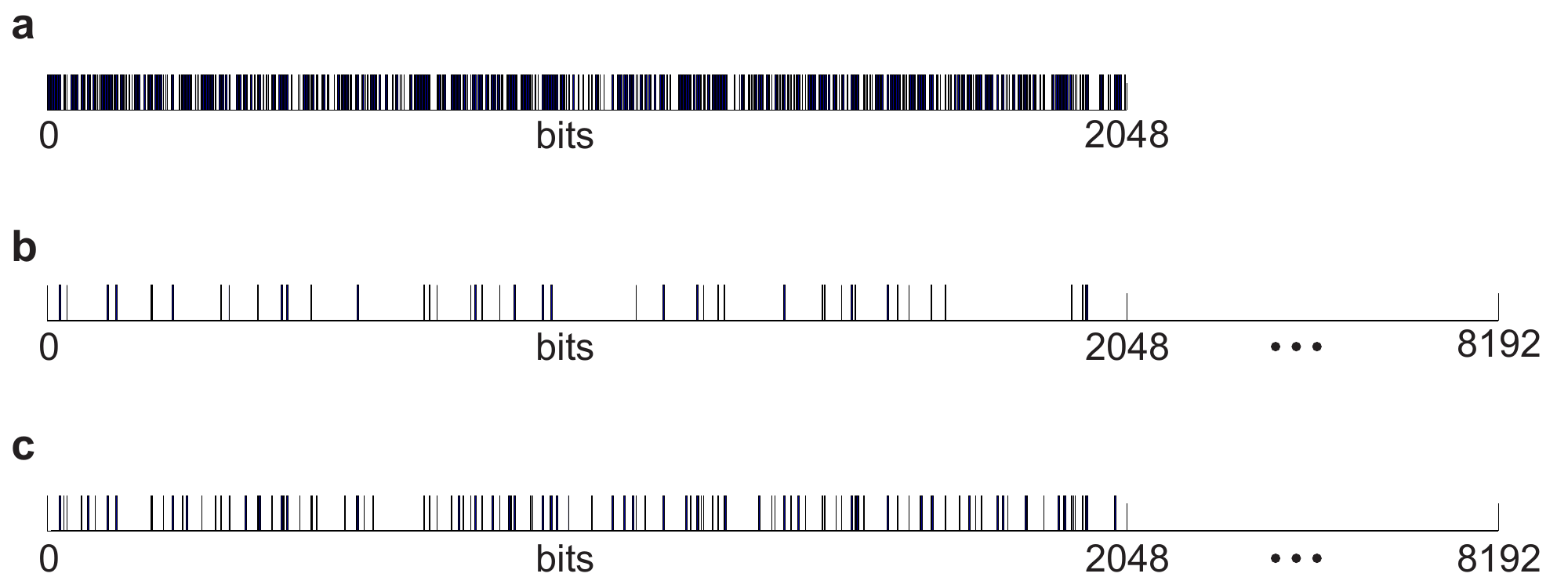} 
		\end{center} 
		\caption{Actions of predetermined complexity. The target complexity of actions was 2048 bits. (a) An action, represented by a completely random bitstring of length 2048 bits. The probability of bits equal to one equals 0.5. Distortions are not allowed. (b) An action, represented by a bitstring of length 8192 bits. The probability of ones equals 0.042. The  string has the complexity close to 2048 bits. Distortions are not allowed. (c) An action, represented by a string of length 8192 bits. The probability of ones equals 0.111. The complexity of the string is about 4096 bits. This action allows for distortions. Other feasible actions can differ from the represented action in no more than 327 bits. Because of the allowed distortions the action has complexity close to 2048 bits as in (a) and (b).}
		\label{AAlg}
	\end{figure}
	\vspace{\baselineskip}

	\section{Discussion}
	In this study, a behavioral act is considered as a set of time series of readings from sensors and actuators of the agent during the behaviors that attains the goal(s) of the task. It is assumed that the behavior, corresponding to the act, may vary from instance to instance. The extent of the variability, along with the length and patterns of behavior are quantified using the Kolmogorov complexity. Thus defined complexity characterizes an apparent behavior and does not take into consideration processes of learning and behavioral control. An algorithm of generating possible descriptions of behavioral acts of predetermined complexity is  reported. Using the algorithm, one can generate abstract sequences of rewards, percepts and actions that would represent successful behavior. Machines could be tested in learning and performing the tasks which are defined by all three components. The work of the algorithm is demonstrated on several examples.
	
	Rewards, percepts and actuators of one and the same behavioral act have their own complexities, though the number of time steps for those time series is obviously the same. The components may have different complexity of typical instances and allowed variability (Fig.\ref{AAlg}). The effects of these factors on behavioral complexity can be taken into account in studies of animals and comparison and testing of artificial agents. In animals, it would be interesting to evaluate complexities of behavioral acts and their relative frequencies to better undestand what the brain does for survival of those animals. 
	
	Comparison of apparent behaviors does not require exactly the same number and characteristics of sensors and actuators in the compared agents. For the agents with smaller dimensions of sensory and actuators spaces extra dimensions can be formally added. Adding zero readings in those dimensions adds only a little to the Kolmogorov complexity.   
	
	It would be also interesting  to evaluate known behavioral tasks for artificial agents, from grabbing objects to playing computer games, to playing Go. For artificial agents all the data, required to evaluate complexity, is available. For example, in the case of neural networks "playing" computer games  \cite{BB13} the time series of rewards are the game points, percepts are the signals from the monitor, and actions are the codes of the joystick. In this example, one of the questions to explore would be about the relation of the network performance to the complexity of the behavior. Intuition suggests that at least some behavioral acts in agents and animals can be compared according to the lexicographic order of the triples $(C(R),C(P),C(A))$,  where $C(R)$, $C(P)$, and $C(A)$ are the complexity of rewards, percepts, and actions respectively. Is it really the case for computer games? Some games can be associated with sets of behavioral acts. For example, in chess a behavioral act is a won game. In such cases the overall complexity of the game can be defined as an average complexity of the corresponding acts. 
	
	A fundamental limitation of the proposed approach is the incomputability of the Kolmogorov complexity \cite{LV08} which is akin to undersampling in the information theoretic approach \cite{P15}. The lengths of compressed files overestimate the complexity. It has to be taken into account.
	
	In the analysis of animal behavior, the approach is limited by the ability of obtaining the readings of sensors and actuators. For simple behaviors, such as reaching for an object by a monkey, only subsets of sensors and actuators can be considered.  However, the most critical readings can already be evaluated relatively well \cite{Sc04}. In the future, more detailed measurements should be possible.  
	
	The proposed approach can be developed in multiple directions. In the examples considered in the text, feasible instances formed sets of simple geometry, such as balls. However, the main formula (\ref{C_A}) does not require that. Sets of the instances can have arbitrary geometry. In particular, they can form  unions of two or more disjoint nonempty subsets. 
	To account for this, one can can consider the ratio of the boundary area to the volume of the sets. 
	
	Another direction is to generate a time series of predetermined complexity using patterns. The method, described in this study, is to set some bits to one randomly. However, even for artificial agents not all combinations of bits are feasible. The approach should be therefore adjusted to using only feasible patterns. Development in this direction is likely related to the studies of grammar complexity (see, e.g. \cite{CS02}).
	
\section*{Acknowledgements} 
The author thanks Dr. Peter Erdi and Dr. Lance Fortnow for valuable discussions.
		
	\medskip
	\small
	\vskip 0.2in
	\bibliography{complexity}

\begin{thebibliography}{15}
\expandafter\ifx\csname natexlab\endcsname\relax\def\natexlab#1{#1}\fi
\providecommand{\url}[1]{\texttt{#1}}
\providecommand{\href}[2]{#2}
\providecommand{\path}[1]{#1}
\providecommand{\DOIprefix}{doi:}
\providecommand{\ArXivprefix}{arXiv:}
\providecommand{\URLprefix}{URL: }
\providecommand{\Pubmedprefix}{pmid:}
\providecommand{\doi}[1]{\href{http://dx.doi.org/#1}{\path{#1}}}
\providecommand{\Pubmed}[1]{\href{pmid:#1}{\path{#1}}}
\providecommand{\bibinfo}[2]{#2}
\ifx\xfnm\undefined \def\xfnm[#1]{\unskip,\space#1}\fi
\bibitem[{Bellemare et~al.(2013)Bellemare, Naddaf, Veness and Bowling}]{BB13}
\bibinfo{author}{Bellemare\xfnm[ M.G.]}, \bibinfo{author}{Naddaf\xfnm[ Y.]},
  \bibinfo{author}{Veness\xfnm[ J.]}, \bibinfo{author}{Bowling\xfnm[ M.]}.
\newblock \bibinfo{title}{The arcade learning environment: An evaluation
  platform for general agents.}
\newblock \bibinfo{journal}{J Artif Intell Res(JAIR)}
  \bibinfo{year}{2013};\bibinfo{volume}{47}:\bibinfo{pages}{253--279}.
\bibitem[{Bialek et~al.(2001)Bialek, Nemenman and Tishby}]{B01}
\bibinfo{author}{Bialek\xfnm[ W.]}, \bibinfo{author}{Nemenman\xfnm[ I.]},
  \bibinfo{author}{Tishby\xfnm[ N.]}.
\newblock \bibinfo{title}{Predictability, complexity, and learning}.
\newblock \bibinfo{journal}{Neural computation}
  \bibinfo{year}{2001};\bibinfo{volume}{13}(\bibinfo{number}{11}):\bibinfo{pages}{2409--2463}.
\bibitem[{Charikar et~al.(2002)Charikar, Lehman, Liu, Panigrahy, Prabhakaran,
  Rasala, Sahai and Shelat}]{CS02}
\bibinfo{author}{Charikar\xfnm[ M.]}, \bibinfo{author}{Lehman\xfnm[ E.]},
  \bibinfo{author}{Liu\xfnm[ D.]}, \bibinfo{author}{Panigrahy\xfnm[ R.]},
  \bibinfo{author}{Prabhakaran\xfnm[ M.]}, \bibinfo{author}{Rasala\xfnm[ A.]},
  \bibinfo{author}{Sahai\xfnm[ A.]}, \bibinfo{author}{Shelat\xfnm[ A.]}.
\newblock \bibinfo{title}{Approximating the smallest grammar: Kolmogorov
  complexity in natural models}.
\newblock In: \bibinfo{booktitle}{Proceedings of the Thiry-fourth Annual ACM
  Symposium on Theory of Computing}. \bibinfo{address}{New York, NY, USA}:
  \bibinfo{publisher}{ACM}; STOC '02; \bibinfo{year}{2002}. p.
  \bibinfo{pages}{792--801}.
\bibitem[{Cilibrasi and Vit{\'a}nyi(2005)}]{CV05}
\bibinfo{author}{Cilibrasi\xfnm[ R.]}, \bibinfo{author}{Vit{\'a}nyi\xfnm[
  P.M.]}.
\newblock \bibinfo{title}{Clustering by compression}.
\newblock \bibinfo{journal}{IEEE Transactions on Information theory}
  \bibinfo{year}{2005};\bibinfo{volume}{51}(\bibinfo{number}{4}):\bibinfo{pages}{1523--1545}.
\bibitem[{Cover and Thomas(1991)}]{CT91}
\bibinfo{author}{Cover\xfnm[ T.]}, \bibinfo{author}{Thomas\xfnm[ J.]}.
\newblock \bibinfo{title}{Elements of Information Theory}.
\newblock \bibinfo{publisher}{John Wiley \& Sons, Inc.}, \bibinfo{year}{1991}.
\bibitem[{Franz and Mallot(2000)}]{FM00}
\bibinfo{author}{Franz\xfnm[ M.O.]}, \bibinfo{author}{Mallot\xfnm[ H.A.]}.
\newblock \bibinfo{title}{Biomimetic robot navigation}.
\newblock \bibinfo{journal}{Robotics and autonomous Systems}
  \bibinfo{year}{2000};\bibinfo{volume}{30}(\bibinfo{number}{1}):\bibinfo{pages}{133--153}.
\bibitem[{Hern{\'a}ndez-Orallo(2015)}]{HO15}
\bibinfo{author}{Hern{\'a}ndez-Orallo\xfnm[ J.]}.
\newblock \bibinfo{title}{On environment difficulty and discriminating power}.
\newblock \bibinfo{journal}{Autonomous Agents and Multi-Agent Systems}
  \bibinfo{year}{2015};\bibinfo{volume}{29}(\bibinfo{number}{3}):\bibinfo{pages}{402--454}.
\bibitem[{Hern{\'e}ndez-Orallo(2000)}]{HO00}
\bibinfo{author}{Hern{\'e}ndez-Orallo\xfnm[ J.]}.
\newblock \bibinfo{title}{Beyond the turing test}.
\newblock \bibinfo{journal}{Journal of Logic, Language and Information}
  \bibinfo{year}{2000};\bibinfo{volume}{9}(\bibinfo{number}{4}):\bibinfo{pages}{447--466}.
\bibitem[{Hutter(2001)}]{Hu01}
\bibinfo{author}{Hutter\xfnm[ M.]}.
\newblock \bibinfo{title}{Towards a universal theory of artificial intelligence
  based on algorithmic probability and sequential decisions}.
\newblock In: \bibinfo{booktitle}{European Conference on Machine Learning}.
  \bibinfo{organization}{Springer}; \bibinfo{year}{2001}. p.
  \bibinfo{pages}{226--238}.
\bibitem[{Legg and Veness(2013)}]{LV13}
\bibinfo{author}{Legg\xfnm[ S.]}, \bibinfo{author}{Veness\xfnm[ J.]}.
\newblock \bibinfo{title}{An approximation of the universal intelligence
  measure}.
\newblock In: \bibinfo{booktitle}{Algorithmic Probability and Friends. Bayesian
  Prediction and Artificial Intelligence}. \bibinfo{publisher}{Springer};
  \bibinfo{year}{2013}. p. \bibinfo{pages}{236--249}.
\bibitem[{Li and Vit{\'a}nyi(2008)}]{LV08}
\bibinfo{author}{Li\xfnm[ M.]}, \bibinfo{author}{Vit{\'a}nyi\xfnm[ P.M.]}.
\newblock \bibinfo{title}{An Introduction to Kolmogorov Complexity and Its
  Applications}.
\newblock \bibinfo{edition}{3rd} ed.
\newblock \bibinfo{publisher}{Springer}, \bibinfo{year}{2008}.
\bibitem[{Palmer et~al.(2015)Palmer, Marre, Berry and Bialek}]{P15}
\bibinfo{author}{Palmer\xfnm[ S.E.]}, \bibinfo{author}{Marre\xfnm[ O.]},
  \bibinfo{author}{Berry\xfnm[ M.J.]}, \bibinfo{author}{Bialek\xfnm[ W.]}.
\newblock \bibinfo{title}{Predictive information in a sensory population}.
\newblock \bibinfo{journal}{Proceedings of the National Academy of Sciences}
  \bibinfo{year}{2015};\bibinfo{volume}{112}(\bibinfo{number}{22}):\bibinfo{pages}{6908--6913}.
\bibitem[{Schmidhuber(2002)}]{Sch02}
\bibinfo{author}{Schmidhuber\xfnm[ J.]}.
\newblock \bibinfo{title}{The speed prior: a new simplicity measure yielding
  near-optimal computable predictions}.
\newblock In: \bibinfo{booktitle}{International Conference on Computational
  Learning Theory}. \bibinfo{organization}{Springer}; \bibinfo{year}{2002}. p.
  \bibinfo{pages}{216--228}.
\bibitem[{Scott(2004)}]{Sc04}
\bibinfo{author}{Scott\xfnm[ S.H.]}.
\newblock \bibinfo{title}{Optimal feedback control and the neural basis of
  volitional motor control}.
\newblock \bibinfo{journal}{Nature Reviews Neuroscience}
  \bibinfo{year}{2004};\bibinfo{volume}{5}(\bibinfo{number}{7}):\bibinfo{pages}{532--546}.
\bibitem[{Tao(2008)}]{TAO08}
\bibinfo{author}{Tao\xfnm[ T.]}.
\newblock \bibinfo{title}{Structure and Randomness: Pages from Year One of a
  Mathematical Blog}.
\newblock Miscellaneous Books. \bibinfo{publisher}{American Mathematical
  Society}, \bibinfo{year}{2008}.

\end{thebibliography}
\end{document}